\documentclass[11pt]{article}

\usepackage[final]{acl}

\usepackage{times}
\usepackage{latexsym}

\usepackage[T1]{fontenc}
\usepackage[utf8]{inputenc}

\usepackage{microtype}

\usepackage{inconsolata}

\usepackage{hyperref}       % hyperlinks
\usepackage{url}            % simple URL typesetting
\usepackage{booktabs}       % professional-quality tables
\usepackage{nicefrac}       % compact symbols for 1/2, etc.
\usepackage{xcolor}         % colors
\usepackage{amsfonts}       % blackboard math symbols
\usepackage{graphicx}
\usepackage{amsmath}
\usepackage{algorithm}
\usepackage{algpseudocode}
\usepackage{mathtools}
\usepackage{enumitem}
\usepackage{tabularx}
\usepackage[most]{tcolorbox}
\tcbuselibrary{skins,raster,listings}
\usepackage{multirow}
\usepackage{float}
\usepackage{wrapfig} % for wraptable

\usepackage{caption}    % for \captionof inside minipage
\usepackage{subcaption}

\usepackage{siunitx}
\usepackage{makecell}
\newcolumntype{Y}{>{\centering\arraybackslash}X}

\title{FedMentor: Domain-Aware Differential Privacy for Heterogeneous Federated LLMs in Mental Health}

\author{
    \textbf{Nobin Sarwar, Shubhashis Roy Dipta}\\
    University of Maryland, Baltimore County\\ 
    \texttt{\{sms2, sroydip1\}@umbc.edu}
}

\begin{document}
\maketitle
\begin{abstract}
Privacy-preserving adaptation of Large Language Models (LLMs) in sensitive domains (e.g., mental health) requires balancing strict confidentiality with model utility and safety. We propose \textbf{FedMentor}, a federated fine-tuning framework that integrates Low-Rank Adaptation (LoRA) and domain-aware Differential Privacy (DP) to meet per-domain privacy budgets while maintaining performance. Each client (domain) applies a custom DP noise scale proportional to its data sensitivity, and the server adaptively reduces noise when utility falls below a threshold. In experiments on three mental health datasets, we show that FedMentor improves safety over standard Federated Learning (FL) without privacy, raising safe output rates by up to three points and lowering toxicity, while maintaining utility (BERTScore F1 and ROUGE-L) within 0.5\% of the non-private baseline and close to the centralized upper bound. The framework scales to backbones with up to 1.7B parameters on single-GPU clients, requiring $<173$\,MB of communication per-round. FedMentor demonstrates a practical approach to privately fine-tune LLMs for safer deployments in healthcare and other sensitive fields.
\end{abstract}

\section{Introduction}

\begin{figure}[!t]
\centering
\includegraphics[width=0.9\columnwidth]{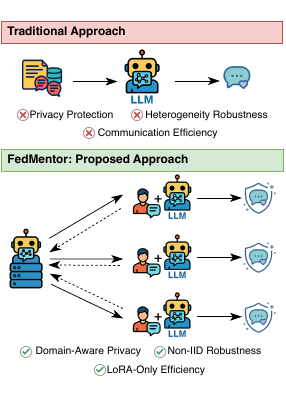}
\caption{Comparison between traditional LLM-based methods for mental health and \textbf{FedMentor}. The baseline lacks privacy protection, robustness to heterogeneity, and communication efficiency, whereas FedMentor introduces domain-aware privacy, achieves robustness under non-IID data, and improves efficiency through LoRA-only updates.}
\label{fig:Fedmentor_treasure}
\vspace{-10pt}
\end{figure}

Mental health arises from interacting cognitive, affective, and behavioral processes that shape individual functioning and societal stability. Demand for scalable support has accelerated interest in LLMs for conversational assistance~\cite{yang2023towards, xu2024mental, yang2024mentallama}. Deployment remains challenging due to strict privacy requirements, limited interpretability, and legal constraints under HIPAA and GDPR~\cite{may2022security, nicholas2020ethics, hipaa1996, gdpr2016}. User inputs may include explicit self harm ideation or other clinical signals that require strong confidentiality (e.g., \emph{[self-harm ideation example]}). These constraints make traditional centralized training challenging for sensitive user data, highlighting the need for techniques that adapt models in a privacy-preserving and communication-efficient manner for healthcare settings. As shown in Figure \ref{fig:Fedmentor_treasure}, our approach addresses these needs by combining domain-aware privacy, robustness under non-IID client distributions, and parameter-efficient updates.

There is growing interest across the AI and healthcare communities in developing trustworthy and scalable mental health chatbots that can provide rapid, accessible, and confidential psychological support. In 2019, mental disorders affected about 970 million people worldwide, or one in eight individuals~\cite{GHDx_2023}. Conversational agents powered by large language models have since emerged as promising tools for supporting mental well-being at scale~\cite{Towards_Healthcare}. The global market for mental health chatbots, valued at \$0.99 billion in 2022, is projected to grow to \$6.51 billion by 2032, underscoring strong clinical and commercial demand~\cite{Towards_Healthcare}. At the same time, deploying LLMs in this setting raises critical challenges in privacy protection, legal compliance, and efficient communication and computation for personalized care~\cite{lai2023psy, hipaa1996, gdpr2016}. These challenges have motivated research into privacy-preserving learning, parameter-efficient adaptation, and federated methods tailored for high-risk domains such as mental health, where user trust and confidentiality are essential.

\begin{figure*}[t]
    \centering
    \includegraphics[width=1.0\textwidth]{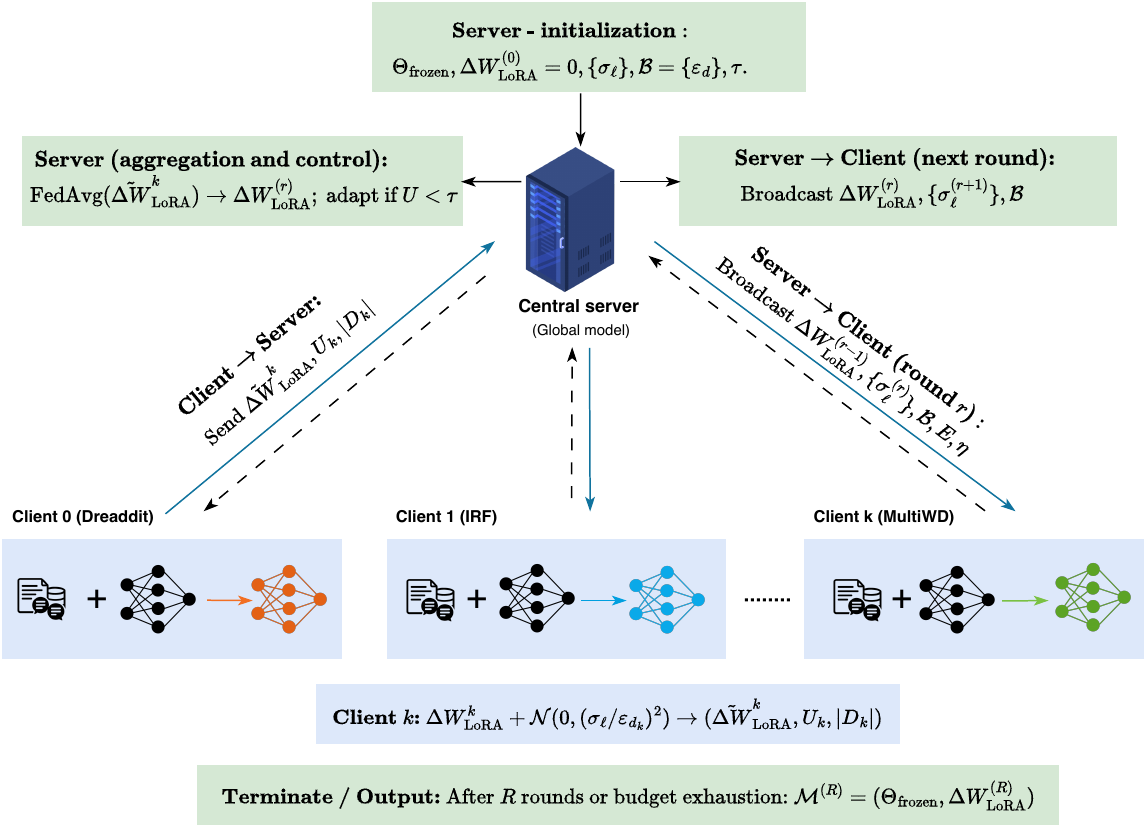}
    \caption{\textbf{FedMentor} pipeline. The server freezes the backbone and initializes LoRA adapters, layer scales, domain privacy budgets, and a utility threshold. Each round it broadcasts the current adapters; clients train LoRA on local data, add Gaussian noise per budget $\varepsilon_d$, and return noised adapters with a utility signal. The server aggregates with FedAvg and reduces noise when utility $<\tau$. After $R$ rounds the model is the frozen backbone plus learned LoRA adapters.}
    \label{fig:fedmentor_architecture}
    \vspace{-10pt}
\end{figure*}

To investigate privacy and communication challenges in adapting LLMs to sensitive mental health domains, we present \textbf{FedMentor}, a federated framework that assigns domain-aware Differential Privacy budgets and fine-tunes only LoRA adapters, aggregated with FedAvg~\cite{mcmahan2017communication}, to achieve private adaptation while preserving utility and fairness under heterogeneity. The overall architecture is shown in Figure~\ref{fig:fedmentor_architecture}. FedMentor is designed for scenarios that demand distinct privacy guarantees across domains and where non-IID data can introduce drift and unequal outcomes. We evaluate the framework on three datasets (Dreaddit~\cite{turcan2019dreaddit}, IRF~\cite{garg2023annotated}, MultiWD~\cite{garg2024multiwd}) and five backbones (MobileLLM-ParetoQ-350M~\cite{liu2024mobilellm}, SmolLM2 (360M and 1.7B)~\cite{allal2025smollm2}, and Qwen3 (0.6B and 1.7B)~\cite{yang2025qwen3}). FedMentor improves safety with minimal loss in utility: for instance, on MultiWD the Toxicity Safe Rate (TSR) rises by two points while toxicity decreases from 2.92 to 1.98 (Table~\ref{tab:MainTable1}), and BERTScore F1~\cite{zhang2020bertscore} and ROUGE-L~\cite{lin2004rouge} remain close to standard FL and near the centralized upper bound. Client-level relevance remains consistent across domains, and efficiency is achieved by communicating only adapter weights through LoRA, keeping computation and bandwidth feasible for single-GPU clients. To the best of our knowledge, this is the first study that combines domain-aware DP~\cite{dwork2014algorithmic,abadi2016deep} with federated LoRA fine-tuning for LLMs in mental health. 

We summarize the main contributions of this work as follows:
\vspace{-5pt}
\begin{itemize}[leftmargin=1.25em]
    \item \textbf{Domain-aware private adaptation for mental health.} We formulate a federated LoRA approach with per-domain $(\epsilon,\delta)$ that protects sensitive mental health text while preserving task utility. FedMentor raises safety (e.g., +2 percentage points TSR with lower toxicity) and keeps BERTScore F1 and ROUGE-L close to FL and near centralized training.
    \vspace{-5pt}
    \item \textbf{Robustness to non-IID heterogeneity.} We demonstrate stable utility and fairness across Dreaddit, IRF, and MultiWD, with per-client relevance closely aligned and no domain collapse. An adaptive noise mechanism maintains performance under strict budgets for sensitive domains.
    \vspace{-5pt}
    \item \textbf{Practical efficiency via adapters.} By exchanging only LoRA adapters, FedMentor reduces communication and memory enough to train backbones up to 1.7B parameters on single-GPU clients, enabling deployment in resource-constrained healthcare settings.
\end{itemize}
\section{FedMentor Foundations: Problem and Notation}

We consider a Federated Learning setting with $K$ distributed clients, where each client $k \in \{1, \ldots, K\}$ holds a private mental health dataset $\mathcal{D}_k$ from domain $d_k \in \{\text{IRF}, \text{Dreaddit}, \text{MultiWD}\}$. The objective is to collaboratively fine-tune a global language model $\mathcal{M}_{\text{global}}$ while ensuring privacy, leveraging domain heterogeneity, and maintaining computational efficiency.

The optimization problem seeks optimal LoRA adapters that minimize the weighted empirical risk:

\vspace{-10pt}

\begin{equation}
\begin{aligned}
\Delta W_{\text{LoRA}}^*
&= \arg\min_{\Delta W_{\text{LoRA}}} \\
&\;\;\sum_{k=1}^{K} \tfrac{|\mathcal{D}_k|}{|\mathcal{D}|}\,
\mathcal{L}_k\!\left(
\Theta_{\text{frozen}} + \Delta W_{\text{LoRA}};\,
\mathcal{D}_k
\right),
\end{aligned}
\end{equation}

where $|\mathcal{D}| = \sum_{j=1}^{K} |\mathcal{D}_j|$ and $\Theta_{\text{frozen}}$ denotes the frozen backbone parameters.

This optimization operates under three critical constraints:

\noindent\textbf{Privacy Constraint.} Each client update must satisfy Differential Privacy (DP)~\cite{dwork2014algorithmic}:

\vspace{-10pt}

\begin{equation}
\label{eq:privacy}
\begin{aligned}
\tilde{\Delta W}_{\text{LoRA}}^{k}
&= \Delta W_{\text{LoRA}}^{k}
  + \mathcal{N}\!\left(0,\left(\frac{\sigma_{l}}{\varepsilon_{d_k}}\right)^{2} I\right),\\
&\qquad \text{satisfying } (\varepsilon_{d_k}, \delta)\text{-DP},
\end{aligned}
\end{equation}

where $\varepsilon_{d_k}\in\{0.5,1.5,2.0\}$ with IRF $=0.5$ (high sensitivity), Dreaddit $=2.0$ (medium), and MultiWD $=1.5$ (low to medium); smaller $\varepsilon$ indicates stronger privacy for more sensitive domains (e.g., interpersonal risk factors).

\vspace{5pt}
\noindent\textbf{Utility and Heterogeneity Constraints.} The global model must maintain clinical viability: 

\vspace{-10pt}

\begin{equation}
\begin{aligned}
\mathcal{U}_m(\mathcal{M}_{\text{global}}) &\geq \tau_m, \\
\forall m &\in 
\bigl\{\text{BERTScore}, \text{SafeRate}, \\
&\quad \text{Relevance}, \text{Perplexity}\bigr\},
\end{aligned}
\end{equation}

where $\tau_m$ represents minimum acceptable thresholds for safe deployment. At the same time, heterogeneity arises from domain shifts:

\vspace{-10pt}

\begin{equation}
\label{eq:heterogeneity}
\begin{aligned}
P_{d_i}(x,y) &\neq P_{d_j}(x,y), \quad \forall i \neq j, \\
d_i, d_j &\in \{\text{IRF}, \text{Dreaddit}, \text{MultiWD}\}.
\end{aligned}
\end{equation}

Instead of treating heterogeneity as a limitation, our formulation leverages complementary distributions to improve robustness and cross-domain generalization. 

\vspace{5pt}

\noindent\textbf{Efficiency Constraint.} Low-Rank Adaptation enables scalable deployment by reducing computation and communication costs:

\vspace{-10pt}

\begin{equation}
\begin{aligned}
\Delta W &= BA,\\
B &\in \mathbb{R}^{d\times r},\quad
A \in \mathbb{R}^{r\times k},\quad
r \ll \min\{d,k\}.
\end{aligned}
\end{equation}

This decomposition reduces trainable parameters from $\mathcal{O}(dk)$ to $\mathcal{O}(r(d+k))$. For communication, if $N$ denotes the total backbone parameters, the transmission cost is reduced by a factor of $\tfrac{r(d_{\text{in}}+d_{\text{out}})}{N} \approx 10^{-3}$, which enables practical deployment on resource-constrained clinical devices while maintaining model expressiveness.
\section{FedMentor Framework}
FedMentor combines Federated Learning (FL)~\cite{mcmahan2017communication}, Low-Rank Adaptation (LoRA)~\cite{hu2021lora}, and domain-aware Differential Privacy (DP)~\cite{dwork2014algorithmic, acharya2020context, noble2022differentially, charles2024fine} to enable scalable, privacy-preserving fine-tuning of LLMs for mental health applications. Unlike traditional FL approaches that transmit full model weights, FedMentor leverages lightweight LoRA updates with client-side noise injection. Algorithm~\ref{alg:fedmentor} outlines the training process.

\subsection{Client Training} 
Each client $k$ maintains dataset $\mathcal{D}_k = \mathcal{D}_k^{\text{train}} \cup \mathcal{D}_k^{\text{val}}$ where $\mathcal{D}_k^{\text{train}} \cap \mathcal{D}_k^{\text{val}} = \emptyset$. During local training, client $k$ receives global LoRA weights $\Delta W_{\text{LoRA}}^{(r-1)}$ and attaches them to the frozen backbone. For each epoch $e \in \{1, \ldots, E\}$ and minibatch $b \subset \mathcal{D}_k^{\text{train}}$, parameters update via:

\begin{equation}
\begin{aligned}
\Delta W_{\text{LoRA}}^k 
&\leftarrow \Delta W_{\text{LoRA}}^k \\
&\hspace{-5pt} - \eta \cdot \nabla_{\Delta W}\,
\mathcal{L}\!\left(
\Theta_{\text{frozen}} + \Delta W_{\text{LoRA}}^k;\,
b
\right)
\end{aligned}
\end{equation}

where $\eta$ is the learning rate and $\mathcal{L}$ is the cross-entropy loss. Notably, only the LoRA weights (typically $<1\%$ of model parameters) are trained, while the backbone remains frozen.

\vspace{5pt}

\noindent\textbf{Client-Side Privacy.}  
To guarantee user-level privacy, clients apply noise before transmission. After local training, for every parameter $p \in \Delta W_k^{(r)}$:
\begin{equation}
\tilde{W}_{k,p}^{(r)} = \Delta W_{k,p}^{(r)} + \mathcal{N}\!\left(0, \left(\frac{\sigma_{l}(p)}{\varepsilon_{d_k}}\right)^{2} I\right),
\end{equation}
where $\sigma_{l}(p)$ is parameter-specific noise scale and $\varepsilon_{d_k}$ follows domain sensitivity from Equation~\ref{eq:privacy}. This implements a Gaussian mechanism with \((\varepsilon_{d(k)}, \delta)\)-Differential Privacy, ensuring all raw weights and local data remain private.

\vspace{5pt}

\noindent\textbf{Layer- and Adapter-Aware Noise Calibration.}
Noise scales are adapted to reflect the relative sensitivity of parameters to perturbation. We classify LoRA weights by their network position, setting $\sigma_{l}(w)=0.01 \cdot \alpha(w)$ for early layers, $\sigma_{l}(w)=0.008 \cdot \alpha(w)$ for middle layers, and $\sigma_{l}(w)=0.005 \cdot \alpha(w)$ for late layers, where $\alpha(w)=1.2$ for LoRA-A matrices and $\alpha(w)=0.8$ for LoRA-B matrices. This scaling reflects empirical observations that early layers and LoRA-A adapters are more sensitive to perturbations, while late layers and LoRA-B adapters are more robust.

\vspace{5pt}

\noindent\textbf{Domain-Specific Privacy Implementation.}
Combining layer-aware noise calibration with domain sensitivity (from Equation~\ref{eq:privacy}), the complete privatized update becomes:
\begin{equation}
\tilde{\Delta W}_{\text{LoRA}}^k = \Delta W_{\text{LoRA}}^k + \mathcal{N}\Big(0, \big(\tfrac{\sigma_{l}(w)}{\varepsilon_{d_k}}\big)^{2} I\Big)
\end{equation}
where $\varepsilon_{d_k}$ follows the domain-specific budgets defined in the privacy constraint.

\vspace{5pt}

\noindent\textbf{Utility-Aware Adjustment.}
To maintain task quality, FedMentor tracks proxy metrics (e.g., BERTScore-F1, safety, relevance). If aggregate utility at round $r$ falls below threshold $\tau$, noise scales are reduced:
\begin{equation}
\sigma_{l}(w) \;\leftarrow\; \alpha \cdot \sigma_{l}(w), \quad \alpha \in (0,1).
\end{equation} 
This privacy–utility feedback loop allows FedMentor to dynamically trade off noise magnitude against task-specific clinical utility.

\subsection{Server Aggregation} 

The server aggregates noised client updates using dataset-weighted FedAvg~\cite{mcmahan2017communication}:

\begin{equation}
\{\tilde{\Delta W}_{\text{LoRA}}^k\}_{k=1}^{K}.
\end{equation}

\vspace{-10pt}

\begin{equation}
\begin{aligned}
\Delta W_{\text{LoRA}}^{(r)} 
&= \sum_{k=1}^{K} \alpha_k \,\tilde{\Delta W}_{\text{LoRA}}^k, \\
\alpha_k 
&= \frac{|\mathcal{D}_k^{\text{train}}|}
{\sum_{j=1}^K |\mathcal{D}_j^{\text{train}}|}
\end{aligned}
\end{equation}

where $\Delta W_{\text{LoRA}}$ represents the collection of all LoRA matrices $\{W_{\text{lora\_A}}^{(l)}, W_{\text{lora\_B}}^{(l)}\}_{l=1}^L$ across $L$ adapted layers, with each matrix aggregated independently.

The global model update combines the frozen quantized backbone with aggregated LoRA weights:

\vspace{-10pt}

\begin{equation}
\mathcal{M}_{\text{global}}^{(r)} 
\;=\; \Theta_{\text{backbone}} \;+\; \Delta W_{\text{LoRA}}^{(r)}.
\end{equation}

\subsection{Communication Efficiency}
FedMentor transmits only LoRA adapter weights rather than full model parameters. The communication cost per client is:
\begin{equation}
\text{CommCost} = \mathcal{O}(r \cdot \sum_l (d_{\text{in}}^{(l)} + d_{\text{out}}^{(l)}))
\end{equation}
where $r$ is the LoRA rank and the sum is over all adapted layers. With typical LoRA configurations ($r \in \{8, 16\}$) applied to models with billions of parameters, this achieves over 99\% compression. For instance, adapting a 1.7B parameter model with rank-16 LoRA requires transmitting only $\sim$2.3 MB versus $\sim$6.8 GB for the full model~\cite{hu2021lora}.

\begin{algorithm}[t]
\caption{\textsc{FedMentor}: Domain-aware DP LoRA for heterogeneous Federated LLMs}
\label{alg:fedmentor}
\small
\textbf{Input:} Datasets $\{\mathcal{D}_k\}_{k=1}^K$, domains $\{d_k\}_{k=1}^K$, rounds $R$, epochs $E$, learning rate $\eta$, privacy budgets $\{\varepsilon_d\}$, thresholds $\{\tau_m\}$\\
\textbf{Output:} Global model $\mathcal{M}^{(R)}=(\Theta_{\text{frozen}}, \Delta W_{\text{LoRA}}^{(R)})$
\begin{algorithmic}[1]
\State \textbf{Server init}
\State $\mathcal{M}^{(0)} \coloneqq (\Theta_{\text{frozen}}, \Delta W_{\text{LoRA}}^{(0)})$
    % \Statex \hfill \Comment{4-bit backbone + LoRA weights}
\State $\sigma_l \coloneqq \text{ClassifyLayers}(\Delta W_{\text{LoRA}}^{(0)})$ 
    % \Statex \hfill \Comment{Layer-importance for LoRA}
\State $\mathcal{B} \coloneqq \{\text{IRF}:0.5,\ \text{Dreaddit}:2.0,\ \text{MultiWD}:1.5\}$ 
    % \Statex \hfill \Comment{Domain budgets}
\For{round $r=1$ to $R$}
    \State \colorbox{pink!20}{\parbox{\dimexpr0.93\linewidth-2\fboxsep\relax}{Broadcast LoRA weights $\Delta W_{\text{LoRA}}^{(r-1)}$ to all clients}} 
        \Statex \hfill \textbf{\Comment{Server update}}
    \For{client $k \in \{1,\ldots,K\}$ \textbf{in parallel}} 
    \Statex \hfill \textbf{\Comment{Client update}}
        \State Attach $\Delta W_{\text{LoRA}}^{(r-1)}$ to frozen $\Theta_{\text{frozen}}$
        \For{epoch $e=1$ to $E$} 
        % \Comment{Train LoRA only}
            \For{batch $b \subset \mathcal{D}_k^{\text{train}}$}
                \State Update:
                \Statex \hspace{70pt}$\Delta W_{\text{LoRA}}^k \leftarrow \Delta W_{\text{LoRA}}^k$
                \Statex \hspace{70pt}$-\ \eta \nabla_{\Delta W}\,\mathcal{L}\!\big(\Theta_{\text{frozen}}+\Delta W_{\text{LoRA}}^k;\, b\big)$
            \EndFor
        \EndFor
        \State $\varepsilon_k \coloneqq \mathcal{B}_{d_k}$ 
        % \Comment{Get domain privacy budget}
        \State Add noise:
        \Statex \hspace{50pt}$\tilde{\Delta W}_{\text{LoRA}}^k \leftarrow 
                \Delta W_{\text{LoRA}}^k + \mathcal{N}\!\big(0,(\sigma_l/\varepsilon_k)^2 I\big)$
        
        \State Send noised LoRA weights $\tilde{\Delta W}_{\text{LoRA}}^k$ to server
    \EndFor
    \State \colorbox{pink!20}{\parbox{\dimexpr0.93\linewidth-2\fboxsep\relax}{Collect $\{\tilde{\Delta W}_{\text{LoRA}}^k\}_{k=1}^K$ from clients}} 
    \Statex \hfill \textbf{\Comment{Client aggregation}}
    \State $\Delta W_{\text{LoRA}}^{(r)} \leftarrow \sum_{k=1}^K\dfrac{|\mathcal{D}_k^{\text{train}}|}
           {\sum_j |\mathcal{D}_j^{\text{train}}|}\;
           \tilde{\Delta W}_{\text{LoRA}}^k$ 
           % \Comment{FedAvg}
    \State If $\mathcal{U}_m < \tau_m$ for any $m$:\; $\sigma_l \leftarrow 0.8\,\sigma_l$ 
    % \Comment{Adapt noise}
    \State Update budget:
    \Statex \hspace{50pt}$\mathcal{B}_d \leftarrow \mathcal{B}_d - 0.1\,\varepsilon_d$ for each domain $d$
\EndFor
\State \Return $\mathcal{M}^{(R)}$ with final LoRA weights $\Delta W_{\text{LoRA}}^{(R)}$
\end{algorithmic}
\end{algorithm}

\section{Experiment Settings}

\textbf{Datasets.} We evaluate FedMentor on three mental health datasets: \textbf{Dreaddit}~\cite{turcan2019dreaddit} for stress detection, \textbf{Interpersonal Risk Factors (IRF)}~\cite{garg2023annotated} for Thwarted Belongingness (TBe) and Perceived Burdensomeness (PBu), and \textbf{MultiWD}~\cite{garg2024multiwd} for multi-label wellness prediction. Together, these corpora cover complementary tasks and form a natural non-IID benchmark across domains. Regulatory requirements under HIPAA and GDPR restrict centralized collection, and lengthy institutional approvals further limit access, producing small, fragmented datasets~\cite{act1996health,regulation2018general,nicholas2020ethics}. Prior FL work mitigates performance degradation from heterogeneity with methods such as SCAFFOLD~\cite{karimireddy2020scaffold}, FedNova~\cite{wang2020tackling}, and Adaptive FedOpt~\cite{reddi2021adaptive}. We adopt a domain-aware FL setting where each dataset defines a client domain and use this setting to evaluate FedMentor under heterogeneous conditions. App.~\ref{app:dataset} provides dataset statistics and preprocessing details.

\vspace{3pt}

\noindent\textbf{Models.} FL places tight limits on client compute, memory, and upload bandwidth; therefore, we adopt compact LLM backbones. We fine-tune five lightweight models from 350M to 1.7B parameters: \textbf{MobileLLM-ParetoQ-350M}~\cite{liu2024mobilellm}, \textbf{SmolLM2} (360M and 1.7B)~\cite{allal2025smollm2}, and \textbf{Qwen3} (0.6B and 1.7B)~\cite{yang2025qwen3}. These backbones integrate quantization and distillation to enhance efficiency, enabling deployment on edge devices and GPUs with limited memory. Their compact size further supports faster client-level fine-tuning and inference, which is critical under the resource and communication constraints of FL~\cite{ro2022scaling, lin2022fednlp, wang2024flora}. To ensure fairness, FedMentor and all baselines are evaluated on identical model architectures so that observed differences arise solely from training strategies. Additional architectural details are provided in App.~\ref{app:models}.

\vspace{3pt}

\noindent\textbf{Baselines.} We compare three setups under identical backbones and LoRA ranks (here, \emph{w/o} denotes \emph{without}). \textbf{(i) Centralized (w/o FL, w/o DP):} pool all datasets and fine-tune a single LoRA adapter on the combined corpus, providing an optimistic upper bound. \textbf{(ii) Federated with LoRA (w/o DP):} run FedAvg with frozen backbones and client-specific LoRA adapters; the server aggregates adapter parameters using data-size weighting to isolate decentralization effects. \textbf{(iii) Federated LoRA under domain-aware DP (FedMentor):} clients allocate domain-specific privacy budgets with layer- and adapter-specific noise scaling; the server performs FedAvg on noised adapters and reduces noise when utility proxies fall below thresholds. Full baseline configurations appear in App.~\ref{app:baselines}.

\begin{table*}[t]
\centering
\scriptsize
\setlength{\tabcolsep}{3pt}      % tighter columns
\renewcommand{\arraystretch}{1.12}
% \resizebox{0.6\textwidth}{!}{
\begin{tabular}{ll
  S[table-format=2.1] S[table-format=2.1]  % SR ZS/FS
  S[table-format=2.2] S[table-format=2.2]  % Tmn ZS/FS
  S[table-format=2.1] S[table-format=2.1]  % B-F1 ZS/FS
  S[table-format=2.2] S[table-format=2.2]  % R-L ZS/FS
  S[table-format=2.1] S[table-format=2.1]  % Rel ZS/FS
}
\toprule
\textbf{Setting} & \textbf{Method} &
\multicolumn{2}{c}{\textbf{SR (\%) $\uparrow$}} &
\multicolumn{2}{c}{\textbf{Tmn (\%) $\downarrow$}} &
\multicolumn{2}{c}{\textbf{B-F1 (\%) $\uparrow$}} &
\multicolumn{2}{c}{\textbf{R-L (\%) $\uparrow$}} &
\multicolumn{2}{c}{\textbf{REL (\%) $\uparrow$}} \\
\cmidrule(lr){3-4}\cmidrule(lr){5-6}\cmidrule(lr){7-8}\cmidrule(lr){9-10}\cmidrule(lr){11-12}
& &
\textbf{ZS} & \textbf{FS} &
\textbf{ZS} & \textbf{FS} &
\textbf{ZS} & \textbf{FS} &
\textbf{ZS} & \textbf{FS} &
\textbf{ZS} & \textbf{FS} \\
\midrule
\multicolumn{12}{c}{\textbf{Dreaddit}}\\
\midrule
\multirow{3}{*}{Central}
& ParetoQ-350M & 98.0 & 98.5 & 1.16 & 0.70 & 83.2 & 86.1 & 6.25 & 13.3 & 23.5 & 18.2 \\
& Qwen3-0.6B & 99.5 & 91.0 & 0.67 & 5.04 & 83.0 & 84.3 & 5.83 & 9.30 & 38.3 & 91.6 \\
& Qwen3-1.7B & 99.5 & 90.0 & 0.49 & 5.07 & 83.4 & 84.3 & 6.38 & 8.83 & 36.0 & 91.3 \\
\cmidrule(lr){1-12}
\multirow{3}{*}{\shortstack{FL\\(w/o DP)}}
& ParetoQ-350M & 99.0 & 99.0 & 0.76 & 0.90 & 82.4 & 82.4 & 3.30 & 3.12 & 22.5 & 22.5 \\
& Qwen3-0.6B & 92.0 & 93.0 & 2.43 & 2.20 & 82.5 & 82.5 & 3.78 & 3.81 & 65.8 & 65.5 \\
& Qwen3-1.7B & 93.0 & 93.0 & 2.35 & 2.52 & 82.4 & 82.4 & 4.03 & 4.07 & 62.9 & 62.4 \\
\cmidrule(lr){1-12}
\multirow{3}{*}{\shortstack{FedMentor\\(FL w/ DP)}}
& ParetoQ-350M & 94.0 & 98.0 & 2.32 & 1.28 & 82.5 & 82.4 & 2.91 & 3.29 & 22.6 & 23.7 \\
& Qwen3-0.6B & 92.0 & 93.0 & 2.23 & 2.19 & 82.5 & 82.5 & 3.82 & 3.78 & 66.3 & 66.1 \\
& Qwen3-1.7B & 93.0 & 92.0 & 2.29 & 2.66 & 82.4 & 82.4 & 4.05 & 4.07 & 62.9 & 62.7 \\
\midrule
\multicolumn{12}{c}{\textbf{IRF}}\\
\midrule
\multirow{3}{*}{Central}
& ParetoQ-350M & 98.0 & 96.0 & 1.04 & 2.21 & 83.2 & 85.3 & 4.89 & 11.9 & 26.7 & 31.0 \\
& Qwen3-0.6B & 92.5 & 66.0 & 4.25 & 17.0 & 83.2 & 83.7 & 4.29 & 6.72 & 43.5 & 92.6 \\
& Qwen3-1.7B & 96.5 & 65.5 & 1.00 & 16.8 & 83.8 & 83.8 & 4.90 & 7.75 & 28.4 & 92.2 \\
\cmidrule(lr){1-12}
\multirow{3}{*}{\shortstack{FL\\(w/o DP)}}
& ParetoQ-350M & 94.0 & 96.0 & 1.27 & 2.52 & 82.5 & 82.5 & 3.36 & 3.50 & 23.9 & 25.4 \\
& Qwen3-0.6B & 78.0 & 78.0 & 8.36 & 8.44 & 82.1 & 82.1 & 3.80 & 3.78 & 70.6 & 71.3 \\
& Qwen3-1.7B & 79.0 & 78.0 & 7.97 & 8.33 & 82.0 & 82.0 & 3.81 & 3.83 & 68.5 & 68.9 \\
\cmidrule(lr){1-12}
\multirow{3}{*}{\shortstack{FedMentor\\(FL w/ DP)}}
& ParetoQ-350M & 95.0 & 92.0 & 1.71 & 3.47 & 82.4 & 82.4 & 3.03 & 3.27 & 25.3 & 24.0 \\
& Qwen3-0.6B & 77.0 & 78.0 & 8.53 & 8.44 & 82.1 & 82.1 & 3.79 & 3.82 & 69.8 & 71.4 \\
& Qwen3-1.7B & 78.0 & 79.0 & 8.37 & 8.18 & 82.0 & 82.0 & 3.82 & 3.85 & 68.0 & 68.2 \\
\midrule
\multicolumn{12}{c}{\textbf{MultiWD}}\\
\midrule
\multirow{3}{*}{Central}
& ParetoQ-350M & 96.0 & 94.5 & 2.25 & 2.78 & 83.0 & 84.6 & 5.00 & 7.06 & 25.5 & 27.0 \\
& Qwen3-0.6B & 95.5 & 66.5 & 2.78 & 17.9 & 84.1 & 82.8 & 5.77 & 5.21 & 27.0 & 88.5 \\
& Qwen3-1.7B & 98.5 & 67.0 & 1.31 & 17.2 & 83.7 & 82.7 & 6.50 & 4.61 & 23.4 & 88.1 \\
\cmidrule(lr){1-12}
\multirow{3}{*}{\shortstack{FL\\(w/o DP)}}
& ParetoQ-350M & 95.0 & 93.0 & 2.26 & 2.92 & 82.4 & 82.3 & 3.57 & 3.39 & 24.3 & 24.2 \\
& Qwen3-0.6B   & 80.0 & 80.0 & 5.91 & 5.86 & 82.0 & 82.0 & 3.34 & 3.38 & 70.9 & 70.2 \\
& Qwen3-1.7B  & 80.0 & 80.0 & 5.57 & 5.66 & 81.9 & 81.9 & 3.43 & 3.42 & 68.3 & 68.3 \\
\cmidrule(lr){1-12}
\multirow{3}{*}{\shortstack{FedMentor\\(FL w/ DP)}}
& ParetoQ-350M & 95.0 & 95.0 & 1.68 & 1.98 & 82.3 & 82.4 & 3.31 & 3.76 & 23.5 & 22.9 \\
& Qwen3-0.6B   & 79.0 & 80.0 & 6.03 & 6.25 & 82.0 & 82.0 & 3.39 & 3.37 & 69.8 & 70.7 \\
& Qwen3-1.7B  & 79.0 & 81.0 & 6.30 & 6.03 & 81.9 & 81.9 & 3.39 & 3.44 & 68.7 & 68.1 \\
\bottomrule
\end{tabular}
% }
\caption{Performance of LLM backbones under different training setups on Dreaddit, IRF, and MultiWD. The table reports Zero-Shot (ZS) and Few-Shot (FS) results for Safe Rate (SR $\equiv$ TSR), Toxicity mean (Tmn), BERTScore F1 (B-F1), ROUGE-L (R-L), and Relevance (REL). Centralized denotes fine-tuning on combined data (upper bound), FL (w/o DP) refers to FL without DP, and FedMentor applies domain-aware DP. In the FL setting, each dataset is treated as a single client to reflect domain heterogeneity. All metrics are in \% (↑ higher is better; ↓ lower is better).}
\label{tab:MainTable1}
\vspace{-10pt}
\end{table*}

\vspace{3pt}
\noindent\textbf{Evaluation Metrics.} We evaluate centralized models with toxicity mean and max, safe rate~\cite{gehman2020realtoxicityprompts}, BERTScore F1~\cite{zhang2020bertscore}, relevance~\cite{reimers2019sentence}, ROUGE-L~\cite{lin2004rouge}, and perplexity~\cite{jelinek1977perplexity}. In federated settings we report toxicity mean, safe rate, BERTScore F1, relevance, and ROUGE-L. With DP we additionally track domain sensitivity, training time, final losses, memory, adapter size, and round-level averages including communication overhead measured by LoRA updates. App.~\ref{app:metrics} provides details on the evaluation metrics.

\begin{figure}[t]
\centering
\begin{minipage}[t]{0.49\columnwidth}
    \centering
    \includegraphics[width=\linewidth]{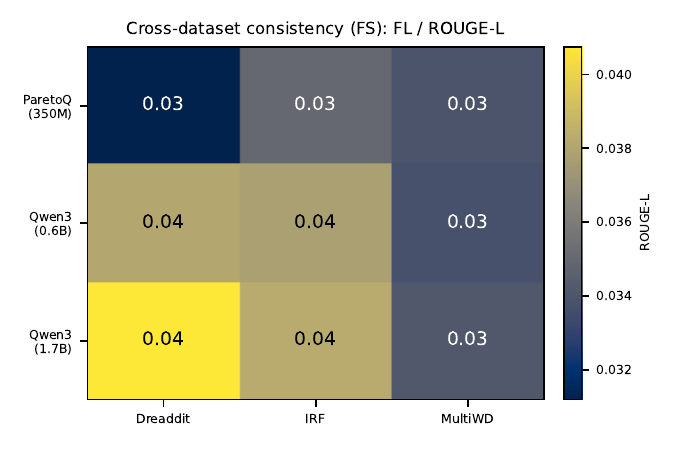}
    \subcaption{FL}
\end{minipage}
\hfill
\begin{minipage}[t]{0.49\columnwidth}
    \centering
    \includegraphics[width=\linewidth]{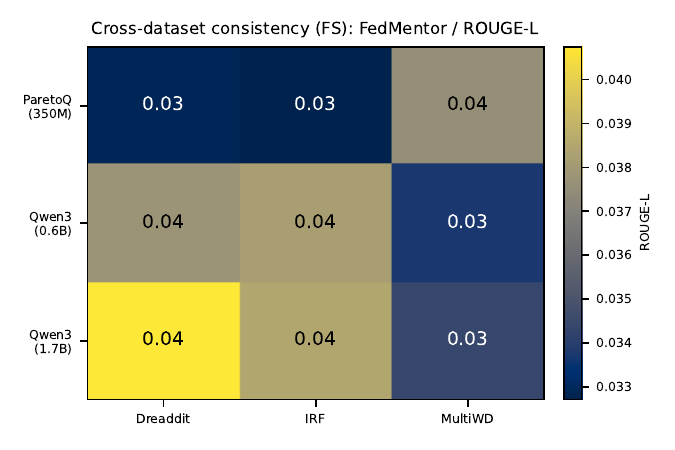}
    \subcaption{FedMentor}
\end{minipage}
\caption{Cross-dataset consistency (FS, ROUGE-L) under FL (left) and FedMentor (right).}
\label{fig:consistency_heatmaps}
\vspace{-10pt}
\end{figure}

\noindent \textbf{Prompt Construction.}
We use two prompting protocols to ensure consistent evaluation across datasets and models. Zero-shot evaluation~\cite{kojima2022large} tests instruction following without the use of in-context examples, relying only on task instructions. Building on this, few-shot evaluation~\cite{brown2020language, sanh2022multitask} performs supervised LoRA adaptation on each client under domain-aware DP with FedAvg aggregation, and inference reuses the same wrapper employed in the zero-shot setting. Full templates, wrappers, and examples appear in App.~\ref{app:prompt}.

\begin{table}[htbp]
\centering
\setlength{\tabcolsep}{3pt}
\renewcommand{\arraystretch}{1.12}
\resizebox{\columnwidth}{!}{
\begin{tabular}{l
  S[table-format=2.2]  % Adpt (MB)
  S[table-format=3.2]  % Comm (MB)
  S[table-format=2.2]  % Mem (GB)
  S[table-format=2.2]  % Time (min)
}
\toprule
\textbf{Model} & \textbf{Adpt (MB)} & \textbf{Comm (MB)} & \textbf{Mem (GB)} & \textbf{Time (min)} \\
\midrule
\multicolumn{5}{c}{\textbf{Global summary (aggregated)}}\\
\midrule
ParetoQ-350M & 16.56 & 49.69  & 33.74 & 7.64 \\
Qwen3-0.6B   & 38.50 & 110.00 & 77.86 & 10.78 \\
Qwen3-1.7B   & 66.50 & 172.90 & 77.86 & 11.37 \\
\midrule
\multicolumn{5}{c}{\textbf{Per-dataset breakdown (Client centric)}}\\
\midrule
\multicolumn{5}{c}{\textbf{Dreaddit}}\\
\midrule
ParetoQ-350M & 16.56 & 49.69  & 33.51 & 4.86 \\
Qwen3-0.6B   & 38.50 & 110.00 & 77.28 & 7.22 \\
Qwen3-1.7B   & 66.50 & 172.90 & 77.86 & 7.80 \\
\midrule
\multicolumn{5}{c}{\textbf{IRF}}\\
\midrule
ParetoQ-350M & 16.56 & 49.69  & 33.62 & 3.96 \\
Qwen3-0.6B   & 38.50 & 110.00 & 77.86 & 5.85 \\
Qwen3-1.7B   & 66.50 & 172.90 & 77.86 & 6.31 \\
\midrule
\multicolumn{5}{c}{\textbf{MultiWD}}\\
\midrule
ParetoQ-350M & 16.56 & 49.69  & 33.74 & 7.64 \\
Qwen3-0.6B   & 38.50 & 110.00 & 76.70 & 11.37 \\
Qwen3-1.7B   & 66.50 & 172.90 & 77.86 & 12.25 \\
\bottomrule
\end{tabular}
}
\caption{Efficiency comparison of LLM backbones under FedMentor. Columns report adapter size (Adpt), communication per-round (Comm), peak GPU memory (Mem), and training time per-round (Time).}
\label{tab:efficiency}
\vspace{-10pt}
\end{table}

\section{Main results}

\textbf{Privacy and Safety.}
FedMentor enforces strict per-domain DP while achieving safer outputs and nearly the same utility as non-private FL (Table~\ref{tab:MainTable1}). Across Dreaddit, IRF, and MultiWD, TSR increases and toxicity decreases, while BERTScore F1 and ROUGE-L change slightly. On MultiWD (few-shot, ParetoQ-350M), TSR improves from 93.0\% to 95.0\% and mean toxicity drops from 2.92 to 1.98. On IRF, the strictest privacy domain, FedMentor achieves 92\% safe outputs versus 96\% for FL and keeps BERTScore F1 and ROUGE-L within 0.1 to 0.2 of the no-DP baseline. Scaling up, FedMentor with Qwen3-0.6B matches the FL safe rate ($\approx$93\%) with identical BERTScore F1, and with Qwen3-1.7B the utility gap narrows further. Overall, FedMentor provides per-domain private training that improves safety over vanilla FL and approaches the centralized upper bound; App.~\ref{app:Supplementaryresults} reports consistent trends for SmolLM2-360M/1.7B.

\begin{figure*}[t]
\centering
\captionsetup{font=small,skip=2pt}

\begin{minipage}[t]{.32\textwidth}\vspace{0pt}\centering
  \includegraphics[width=\linewidth,trim=0 6 0 0,clip]{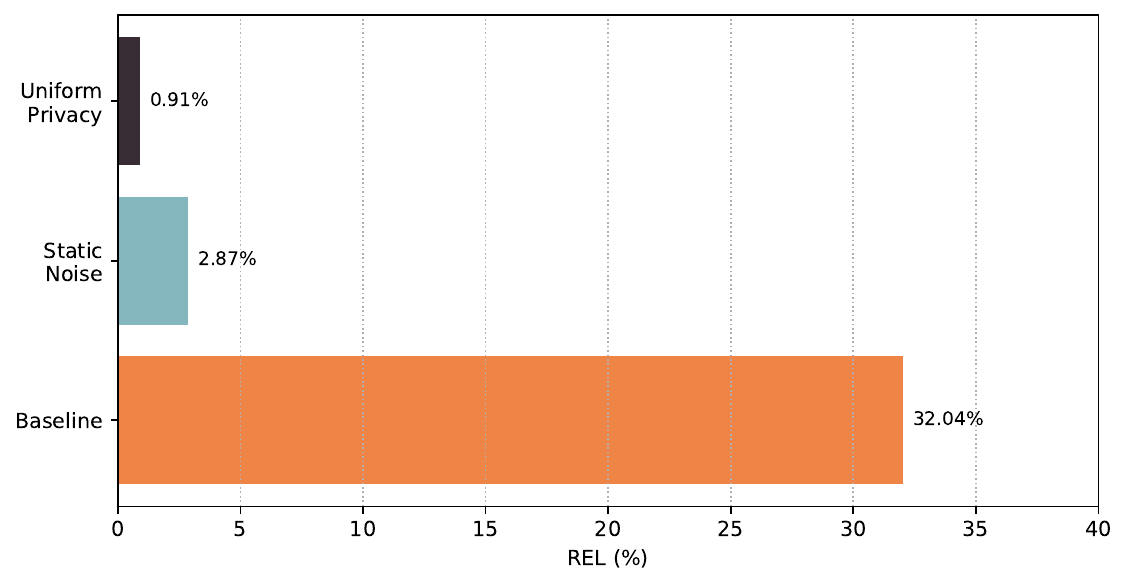}
\end{minipage}\hfill
\begin{minipage}[t]{.32\textwidth}\vspace{0pt}\centering
  \includegraphics[width=0.97\linewidth,trim=0 6 0 0,clip]{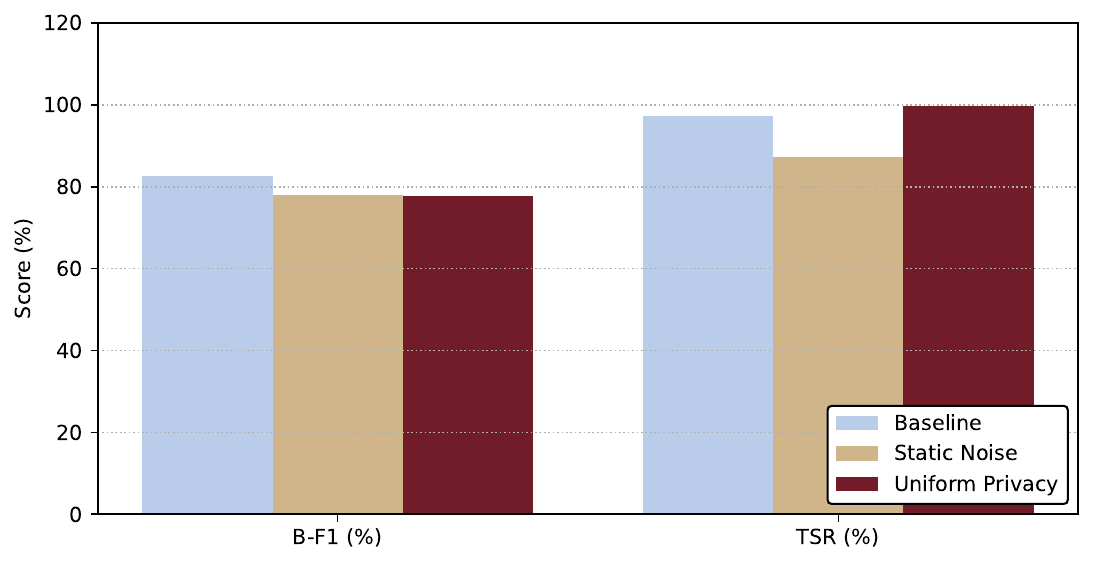}
\end{minipage}\hfill
\begin{minipage}[t]{.32\textwidth}\vspace{0pt}\centering
  \includegraphics[width=0.89\linewidth,trim=0 6 0 0,clip]{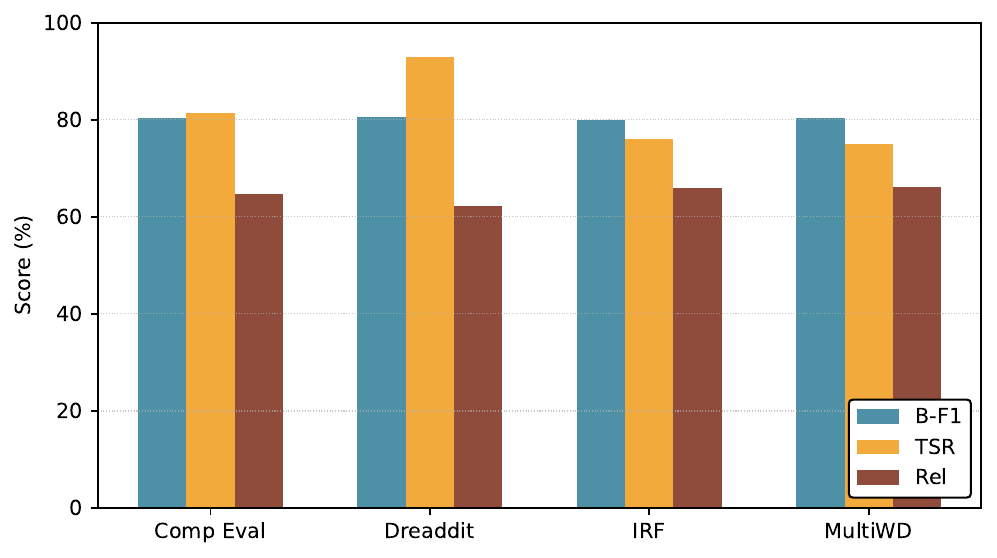}
\end{minipage}

\vspace{2pt}
\caption{ParetoQ-350M: (a) REL and (b) B-F1 with TSR under Baseline, Static, and Uniform. Qwen3-1.7B: (c) $\varepsilon$ ablation of B-F1, TSR, and REL on global evaluation and on Dreaddit, IRF, and MultiWD. All panels report the final global model after 8 rounds (2 local epochs per-round).}
\label{fig:rel_bf1_tsr_paretoq350m_qwen3_1p7b_eps}
\end{figure*}

\vspace{3pt}

\noindent\textbf{Utility under Heterogeneity.}
FedMentor sustains competitive utility on non-IID data, and few-shot fine-tuning consistently outperforms zero-shot prompting. In Table~\ref{tab:MainTable1}, under few-shot training FedMentor’s performance closely matches FL without DP: for example, on Dreaddit the FedMentor BERTScore F1 and ROUGE-L are within 0.1 and 0.2 of the no-DP FL values, and safe rate remains high (98\% vs 99\%). Few-shot adaptation yields clear gains over zero-shot across all models. In particular, relevance and BERTScore improve by roughly 1–3 points with few-shot training compared to zero-shot (e.g., IRF relevance rises from 69.8\% to 71.4\% under FedMentor with Qwen3-0.6B). We observe no domain collapse: Figure~\ref{fig:consistency_heatmaps} shows only modest variation across datasets. Furthermore, at the client level FedMentor exhibits consistent behavior with FL: Figure~\ref{fig:client_level_rel} (App.~\ref{app:Supplementaryresults}) shows per-client relevance scores overlapping (within $\pm$1\%), indicating uniform utility across clients. Although absolute metrics are lower without fine-tuning, the relative gap between FedMentor and FL remains stable, reinforcing that strong privacy comes at minimal utility cost under heterogeneous data.

\begin{table*}[t]
\centering
% \scriptsize
\setlength{\tabcolsep}{3pt}
\renewcommand{\arraystretch}{1.12}
\sisetup{table-number-alignment=center}
\resizebox{\textwidth}{!}{
\begin{tabular}{@{}l l c l l l
  S[table-format=2.2]
  S[table-format=3.0]
  S[table-format=3.0]
  S[table-format=2.2]
  S[table-format=2.0]
@{}}
\toprule
\textbf{Privacy Strategy} &
\textbf{Target Scope} &
$\boldsymbol{\varepsilon_\text{glob}}$ &
\textbf{Budgets (D/I/M)} &
$\boldsymbol{\sigma}$ &
$\boldsymbol{\tau}$ &
\textbf{TSR mean} &
\textbf{TSR min} &
\textbf{TSR max} &
\textbf{TSR std} &
\textbf{Spread} \\
\midrule
Baseline & \makecell[l]{Domain\\Spec.}
& \multicolumn{1}{c}{--}
& \textbf{2.0 / 0.5 / 1.5}
& derived & \multicolumn{1}{c}{--}
& 97.33 & 94 & 100 & 2.49 & 6 \\

\multirow{3}{*}{IRF Sensitivity} & \multirow{3}{*}{\makecell[l]{Domain\\Spec.}}
& fixed & \textbf{2.0 / 0.1 / 1.5} & derived & \multicolumn{1}{c}{--} & 78.33 & 66 & 88  & 9.18 & 22 \\
& & fixed & \textbf{2.0 / 0.5 / 1.5} & derived & \multicolumn{1}{c}{--} & 75.33 & 57 & 90  & 13.72 & 33 \\
& & fixed & \textbf{2.0 / 1.0 / 1.5} & derived & \multicolumn{1}{c}{--} & 77.33 & 70 & 86  & 6.60 & 16 \\

Static Noise & Global
& \multicolumn{1}{c}{--}
& \multicolumn{1}{c}{--}
& $\sigma{=}0.008$
& \multicolumn{1}{c}{--}
& 87.33 & 80 & 98 & 7.72 & 18 \\

Uniform Privacy & Global
& 1.0
& \multicolumn{1}{c}{--}
& derived
& \multicolumn{1}{c}{--}
& 99.67 & 99 & 100 & 0.47 & 1 \\

Utility Threshold & Global
& \multicolumn{1}{c}{--}
& \multicolumn{1}{c}{--}
& \multicolumn{1}{c}{--}
& \makecell[l]{B-F1\\$\tau{=}0$}
& 74.67 & 65 & 85 & 8.18 & 20 \\
\bottomrule
\end{tabular}
}
\caption{Client fairness after 8 federated rounds (ParetoQ-350M). Budgets are reported as D/I/M where D = Dreaddit, I = IRF, and M = MultiWD. IRF Sensitivity varies only the IRF budget; Dreaddit and MultiWD are fixed at 2.0 and 1.5. Metrics in \% (↑ higher is better; ↓ lower is better).}
\label{tab:privacy_strategies_ParetoQ}
\vspace{-5pt}
\end{table*}

\vspace{2pt}

\noindent\textbf{Practical Efficiency.}
FedMentor satisfies practical efficiency requirements in communication, memory, and speed on single-GPU clients. Table~\ref{tab:efficiency} reports compact LoRA updates: 16.56\,MB for ParetoQ-350M, 38.50\,MB for Qwen3-0.6B, and 66.50\,MB for Qwen3-1.7B. With three clients, this translates to about 49.7\,MB, 110.0\,MB, and 172.9\,MB communicated per-round, since only adapters, not full weights, are exchanged. Peak memory per client remains about 33.7\,GB for the 350M model and 77.9\,GB for the Qwen3 models, within a single 80\,GB GPU; per-round Qwen3-1.7B memory for the IRF $\varepsilon$ sweep in Table~\ref{tab:client_memory_gb_Qwen3_1.7B_IRF} (App.~\ref{app:Supplementaryresults}) shows all clients below the device limit. Average round time (two local epochs, three clients) is 7.6, 10.8, and 11.4 minutes for the 350M, 0.6B, and 1.7B backbones. Dataset trends follow size: IRF is fastest (3.96\,min at 350M), Dreaddit is intermediate, and MultiWD is slowest (up to 12.3\,min at 1.7B). Communication cost is constant across datasets (50–173\,MB) as it depends on model size only. Together with Table~\ref{tab:MainTable1}, these results indicate low overhead and practical training times.
\section{Ablation Studies}

\textbf{Uniform Privacy.}
Uniform Privacy applies a single global privacy budget to all domains. In Figure~\ref{fig:rel_bf1_tsr_paretoq350m_qwen3_1p7b_eps}a-b and Table~\ref{tab:privacy_strategies_ParetoQ}, this choice markedly improves fairness in safety: with a uniform $\epsilon{=}1.0$, the TSR concentrates at $\sim$99-100\% and the client spread contracts to 1\% (vs.\ 6\% under the domain-specific baseline). In contrast, the static noise variant shows lower average safety (87.33\% TSR) and a larger spread (18\%). The utility-gated variant with $\tau{=}0$ yields the lowest safety (74.67\% TSR). BERTScore F1 and REL remain close to the baseline across settings, indicating that the fairness gains from a uniform budget come with only modest utility change. All panels in Figure~\ref{fig:rel_bf1_tsr_paretoq350m_qwen3_1p7b_eps} report the final global model after 8 rounds.

\vspace{3pt}

\noindent\textbf{Static Noise.} We ablate adaptive noise by fixing the noise scale across rounds. Table~\ref{tab:privacy_strategies_ParetoQ} shows that Static Noise reduces safety and increases disparity: TSR drops to 87.33\% and the client spread widens to 18\% (baseline 97.33\%, spread 6\%). The rightward bars in Figure~\ref{fig:rel_bf1_tsr_paretoq350m_qwen3_1p7b_eps}a–b align with this trend, reflecting weaker safety and utility than the domain-aware and uniform settings. The result indicates that dynamic privacy control is beneficial for cross-domain balance, whereas a fixed noise schedule amplifies inter-client gaps without yielding compensatory gains.

\vspace{3pt}

% App.~\ref{app:irf_sweep} covers the IRF \texorpdfstring{$\boldsymbol{\epsilon}$}{epsilon} sweep; App.~\ref{app:utility_threshold} the utility-threshold ablation; App.~\ref{app:memory_scaling} per-client memory scaling across rounds.

% \vspace{3pt}

% \subsection{IRF \texorpdfstring{$\boldsymbol{\epsilon}$}{epsilon} Sweep.} 
% \label{app:irf_sweep}

\noindent\textbf{IRF \texorpdfstring{$\boldsymbol{\epsilon}$}{epsilon} Sweep.} We vary only the IRF domain budget while holding Dreaddit and MultiWD fixed (Figure~\ref{fig:rel_bf1_tsr_paretoq350m_qwen3_1p7b_eps}c; Table~\ref{tab:irf-eps-seep}). For Qwen3-1.7B, tightening privacy from $\epsilon{=}1.0$ to $\epsilon{=}0.1$ increases safety (TSR $81.33\%\!\rightarrow\!92.00\%$) with a small B-F1 change ($80.33\%\!\rightarrow\!82.40\%$) and a modest REL shift ($64.78\%\!\rightarrow\!62.70\%$). For ParetoQ-350M, REL remains low across the sweep (about 5\%), and TSR varies between 75.33\% and 78.33\%. These trends indicate that stronger IRF privacy can raise safety for larger models with limited utility loss, whereas smaller backbones display nearly constant utility under the same adjustments.

\begin{table}[t]
\centering
\small
\setlength{\tabcolsep}{2.5pt}
\renewcommand{\arraystretch}{1.0}
\begin{tabular}{@{}l S[table-format=2.1] S[table-format=2.1] S[table-format=2.1]@{}}
\toprule
\textbf{Method} & {\textbf{TSR}} & {\textbf{REL}} & {\textbf{B-F1}} \\
\midrule
\multicolumn{4}{c}{$\boldsymbol{\varepsilon = 0.1}$}\\
\midrule
ParetoQ-350M   & 78.3 &  5.2 & 78.0 \\
Qwen3-0.6B     & 83.7 & 69.4 & 82.2 \\
Qwen3-1.7B     & 92.0 & 62.7 & 82.4 \\
\midrule
\multicolumn{4}{c}{$\boldsymbol{\varepsilon = 0.5}$}\\
\midrule
ParetoQ-350M   & 75.3 &  4.9 & 78.1 \\
Qwen3-0.6B     & 83.7 & 69.4 & 82.2 \\
Qwen3-1.7B     & 92.0 & 62.7 & 82.4 \\
\midrule
\multicolumn{4}{c}{$\boldsymbol{\varepsilon = 1.0}$}\\
\midrule
ParetoQ-350M   & 77.3 &  5.1 & 78.0 \\
Qwen3-0.6B     & 83.7 & 69.4 & 82.2 \\
Qwen3-1.7B     & 81.3 & 64.8 & 80.3 \\
\bottomrule
\end{tabular}
\caption{IRF $\varepsilon$ sweep with other DP hyperparameters fixed. Global summary of Toxicity Safe Rate (TSR), Relevance (REL), and BERTScore F1 (B-F1) after 8 federated rounds across 3 datasets. Metrics in \% (↑ higher is better; ↓ lower is better).}
\label{tab:irf-eps-seep}
\vspace{-5pt}
\end{table}

\vspace{3pt}
% \subsection{Utility Threshold Calibration (\texorpdfstring{$\tau$}{tau}).} 
% \label{app:utility_threshold}

\noindent\textbf{Utility Threshold Calibration (\texorpdfstring{$\tau$}{tau}).} We evaluate a utility-driven noise adjustment where the server reduces noise only if B-F1 falls below a threshold $\tau$. Table~\ref{tab:privacy_strategies_ParetoQ} shows that the $\tau{=}0$ setting yields the lowest safety, with TSR at 74.67\% and a spread of 20\%, compared to the baseline TSR of 97.33\% and spread of 6\%. The observation is consistent with the ParetoQ-350M trends in Figure~\ref{fig:rel_bf1_tsr_paretoq350m_qwen3_1p7b_eps}a–b: aggressive utility gating undermines safety, indicating the need for a calibrated threshold.

% \subsection{Per-Client Memory Scaling across Rounds.} 
% \label{app:memory_scaling}

\noindent\textbf{Per-Client Memory Scaling across Rounds.} Table~\ref{tab:client_memory_gb_Qwen3_1.7B_IRF} reports memory usage per client for Qwen3-1.7B under the IRF $\epsilon{=}1.0$ setting. Memory increases smoothly over rounds: at round 0, clients use about 18.85, 20.04, and 21.23 GB; by round 7, these reach 43.78, 44.97, and 46.16 GB. The monotonic rise indicates predictable scaling during training while keeping per-client usage within a narrow band across clients.

\begin{table}[t]
\centering
\setlength{\tabcolsep}{3pt}
\renewcommand{\arraystretch}{1.12}
\sisetup{table-number-alignment=center}
\resizebox{\columnwidth}{!}{
\begin{tabular}{@{} c
  S[table-format=2.2]
  S[table-format=2.2]
  S[table-format=2.2] @{}}
\toprule
\textbf{Round} &
\textbf{Client 0 (GB)} &
\textbf{Client 1 (GB)} &
\textbf{Client 2 (GB)} \\
\midrule
0 & 18.85 & 20.04 & 21.23 \\
1 & 22.42 & 23.60 & 24.79 \\
2 & 25.98 & 27.16 & 28.35 \\
3 & 29.54 & 30.72 & 31.91 \\
4 & 33.10 & 34.29 & 35.47 \\
5 & 36.66 & 37.85 & 39.03 \\
6 & 40.22 & 41.41 & 42.59 \\
7 & 43.78 & 44.97 & 46.16 \\
\bottomrule
\end{tabular}
}
\caption{Per-round client memory usage for Qwen3-1.7B under the IRF $\epsilon$-sensitivity test with $\epsilon=1.0$. 
The sweep varies only the IRF domain privacy budget, while Dreaddit and MultiWD remain fixed at 2.0 and 1.5, respectively.}
\label{tab:client_memory_gb_Qwen3_1.7B_IRF}
\end{table}

\section{Related Works} 
\label{RelatedWorks}

\textbf{LLMs in Mental Health.} LLMs are rapidly adopted for mental health applications, supporting tasks such as condition detection, diagnosis, and therapeutic dialogue~\cite{wang2021evaluation, yang2023mentalllama, yang2023towards, xu2024mental, wang2024patient, mohammadi2024welldunn, zheng2025promind, chan2025prompt, hengle2024still, haider2025mental}. While these systems demonstrate strong potential for supportive interaction, most current approaches are centralized and depend on aggregating sensitive dialogue data. Such data are scarce due to confidentiality concerns, and strict regulations, including HIPAA and GDPR, further constrain sharing~\cite{kwesi2025exploring, baidal2025guardians, nguyen2025large}. As a result, datasets are often small and biased, which limits the robustness and generalization of models.

These limitations reveal a critical privacy-utility trade-off that existing LLM methods rarely address, reducing their feasibility in real-world clinical contexts~\cite{lin2023towards, chung2023challenges, gabriel2024can, kumar2024unlocking}. FedMentor addresses this challenge by integrating Federated Learning with domain-aware Differential Privacy, enabling collaborative training without centralizing data. This framework allows models to learn from sensitive mental health texts while preserving confidentiality, thereby enhancing both trust and clinical applicability.

\textbf{FL for Mental Health.} Federated Learning (FL) has demonstrated performance comparable to centralized training while satisfying strict privacy regulations such as HIPAA and GDPR~\cite{lieftink2024potential, peng2024depth, woisetschlager2024federated}. By training directly on decentralized data, FL is well-suited for mobile health and multi-clinic settings. In mental health, FedTherapist~\cite{shin2023fedtherapist} leverages on-device FL with speech and keyboard signals to track depression, stress, and mood, and FedMood~\cite{xu2021fedmood} introduces a multi-view framework for depression diagnosis using heterogeneous mobile health data. These studies, along with work on mobile sensing, electronic health records, and multi-institutional collaboration, demonstrate the potential of FL for sensitive mental health applications~\cite{rauniyar2023federated}.

Building on these foundations, recent work has extended FL to LLMs for conversational and diagnostic support. FedMentalCare~\cite{sarwar2025fedmentalcare} combines FL with LoRA to fine tune lightweight LLMs while reducing communication costs. However, many methods assume homogeneous data and only partly address scalability and efficiency. We introduce FedMentor, which enables heterogeneous FL through explicit domain modeling, LoRA-based communication reduction, and domain-aware DP for practical and privacy-preserving LLM fine tuning in mental health.

App.~\ref{app:related} discusses related work on Differential Privacy in FL.

\section{Conclusion}
We introduced FedMentor, a federated fine-tuning framework that pairs domain-aware Differential Privacy with LoRA adapters for sensitive mental health dialogue. Across heterogeneous domains, FedMentor raises safety while keeping utility near non-private Federated Learning and near centralized training. Practicality follows from communicating adapter weights only, which keeps memory and bandwidth within single GPU budgets. Ablations confirm that domain-specific budgets and adaptive noise are essential, since removing either reduces accuracy and amplifies client disparities. The study utilizes a limited set of datasets and automated metrics without clinician review. Future work will scale federations, incorporate clinician-in-the-loop approaches, and conduct multilingual evaluation, as well as extend fairness analyses.

\section*{Limitations}

Several factors limit the present study. First, the evaluation utilizes three English social-media datasets, treating each dataset as a single client, which limits external validity for larger cross-silo federations and clinical conversations. Second, assessment relies on automatic quality and safety proxies, including detector-based toxicity and embedding similarity, without clinician review or user studies. Third, the work reports no audit of demographic or linguistic subgroups and does not provide round-level composition accounting or attack-based testing for privacy loss. Fourth, system results reflect an A100 80 GB environment and a small number of clients, so hardware and scale shifts may alter efficiency. Finally, comparisons do not include alternative differentially private fine-tuning strategies beyond LoRA-based adapters.

\bibliography{custom}

% \clearpage
\appendix

\section{Additional Related Works}
\label{app:related}

\textbf{Differential Privacy in FL.} Differential privacy (DP) has become a standard approach for protecting sensitive data in ML, including FL~\cite{dwork2014algorithmic, abadi2016deep, girgis2021shuffled, noble2022differentially}. Applying DP in practice, however, remains difficult, particularly in mental health settings. Traditional methods that uniformly clip gradients and add noise to updates often degrade performance in healthcare, where datasets are small and signals are fine-grained~\cite{mcmahan2018learning}. Although DP has been studied in areas such as electronic health records and medical imaging, its application in mental health remains limited, with only a few recent studies exploring privacy-by-design strategies~\cite{choudhury2019differential,  sawhney2023much, romijnders2024protect}. Recent research further highlights that not all model components or data are equally sensitive, motivating personalized and domain-aware DP strategies that adapt privacy to data characteristics~\cite{ebadi2015differential, jorgensen2015conservative, liu2021projected, liu2022privacy, liu2024cross}. In LLMs, parameter-efficient methods such as DP-LoRA~\cite{liu2025differentially} reduce risk and communication costs by perturbing only low-rank adapter updates. However, most approaches still adopt a one-size-fits-all noise and overlook the unique sensitivities of mental health text. To overcome this limitation, we introduce FedMentor, a framework that integrates domain-aware DP with heterogeneous FL to provide strong privacy guarantees while preserving clinical utility.

\begin{figure}[t]
\centering
\begin{tcolorbox}[
  enhanced,
  colback=white,
  colframe=black,
  title={Dataset-Specific Instruction Prompts},
  fonttitle=\bfseries,
  fontupper=\small,
]
\textbf{Dreaddit.}  
Analyze this Reddit post for signs of psychological stress. Provide a brief, supportive response in 1–2 sentences. If stress is present, acknowledge feelings and suggest one helpful action. If no stress, provide encouragement. Be empathetic but avoid medical advice.\\
\textit{Post:} \{post\}

\medskip
\textbf{IRF.}  
Analyze this post for signs of interpersonal risk (thwarted belongingness or perceived burdensomeness). Provide a brief, supportive response in 1–2 sentences. Be empathetic but avoid medical advice.\\
\textit{Post:} \{post\}

\medskip
\textbf{MultiWD.}  
Analyze this post for social and emotional wellness concerns. Social wellness involves relationships, connection, and community. Emotional wellness involves managing feelings, stress, and mental health. Provide a brief, supportive response in 1–2 sentences. Be empathetic and suggest practical coping strategies.\\
\textit{Post:} \{post\}
\end{tcolorbox}
\caption{Dataset-specific instructions}
\label{fig:dataset_specific_instructions}
\vspace{-10pt}
\end{figure}

\begin{figure}[t]
\centering
\begin{tcolorbox}[
  enhanced,
  width=\columnwidth,
  colback=white,
  colframe=black,
  title={Zero-Shot Model Wrappers},
  fonttitle=\bfseries,
  fontupper=\footnotesize,
]
\textbf{Qwen}
\begin{verbatim}
<|im_start|>user
{INSTRUCTION}
Post: {post}
<|im_end|>
<|im_start|>assistant
\end{verbatim}

\vspace{5pt}
\textbf{SmolLM2}
\begin{verbatim}
### Input:
{INSTRUCTION}
Post: {post}

### Response:
\end{verbatim}

\vspace{5pt}
\textbf{MobileLLM}
\begin{verbatim}
{INSTRUCTION}
Post: {post}

Response:
\end{verbatim}
\end{tcolorbox}
\caption{Prompt templates: Zero-shot wrappers for \textbf{Qwen}, \textbf{SmolLM2}, and \textbf{MobileLLM}.}
\label{fig:prompt_templates_zeroshot}
\end{figure}

\begin{figure}[t]
\centering
\begin{tcolorbox}[
  enhanced,
  width=\columnwidth,
  colback=white,
  colframe=black,
  title={Few-Shot and Train-Time Shape},
  fonttitle=\bfseries,
  fontupper=\footnotesize,
]
\textbf{Few-shot wrapper (generic $k$).}
\begin{verbatim}
[template prefix]
{INSTRUCTION}

Example 1
Post: {post_1}
Response: {response_1}
...
Example k
Post: {post_k}
Response: {response_k}

Post: {post}
Response:
[template suffix]
\end{verbatim}

\textbf{Exact train-time prompt shape.}
\begingroup
  \hfuzz=3pt
  \overfullrule=0pt
\begin{verbatim}
[template prefix] +
{INSTRUCTION} + "\n\nPost: {post}" +
[template suffix] +
{supervised target response} +
[template end]
\end{verbatim}
\endgroup
\end{tcolorbox}
\caption{Prompt templates: Few-shot wrapper and exact train-time prompt shape.}
\label{fig:prompt_templates_fewshot}
\end{figure}

\section{Additional Experiment Settings}
\subsection{Dataset Details}
\label{app:dataset}

\noindent\textbf{Dreaddit}~\cite{turcan2019dreaddit} is a Reddit corpus for stress detection with 3,553 text segments labeled as stressed or not stressed. We use the official splits, add a 10\% stratified validation set, and frame the task as supportive-response generation with stress-sensitive instructions. Labels are retained for monitoring and prompt-target selection.

\medskip

\noindent\textbf{IRF}~\cite{garg2023annotated} contains 3,522 posts annotated for Thwarted Belongingness and Perceived Burdensomeness. We follow the original splits and map the multi-label annotations to a binary indicator of interpersonal risk, aligning with our generation objective. Labels are used for utility tracking and evaluation.

\medskip

\noindent\textbf{MultiWD}~\cite{garg2024multiwd} provides 3,281 Reddit posts annotated with six wellness dimensions. We form a 10\% validation set, apply minority upsampling to balance classes, and reduce the labels to a binary indicator for supportive-response monitoring while retaining per-dimension information for conditioning and evaluation. 

\medskip

\noindent Each dataset is assigned to a distinct client, forming a natural non-IID configuration with domain-specific shifts. Preprocessing removes empty rows, standardizes text fields, and coerces labels to ${0,1}$, supporting a uniform generation-centric framing across stress, interpersonal risk, and wellness.

\begin{figure*}[htbp]
  \centering
  \begin{subfigure}[t]{0.33\textwidth}
    \centering
    \includegraphics[width=\linewidth]{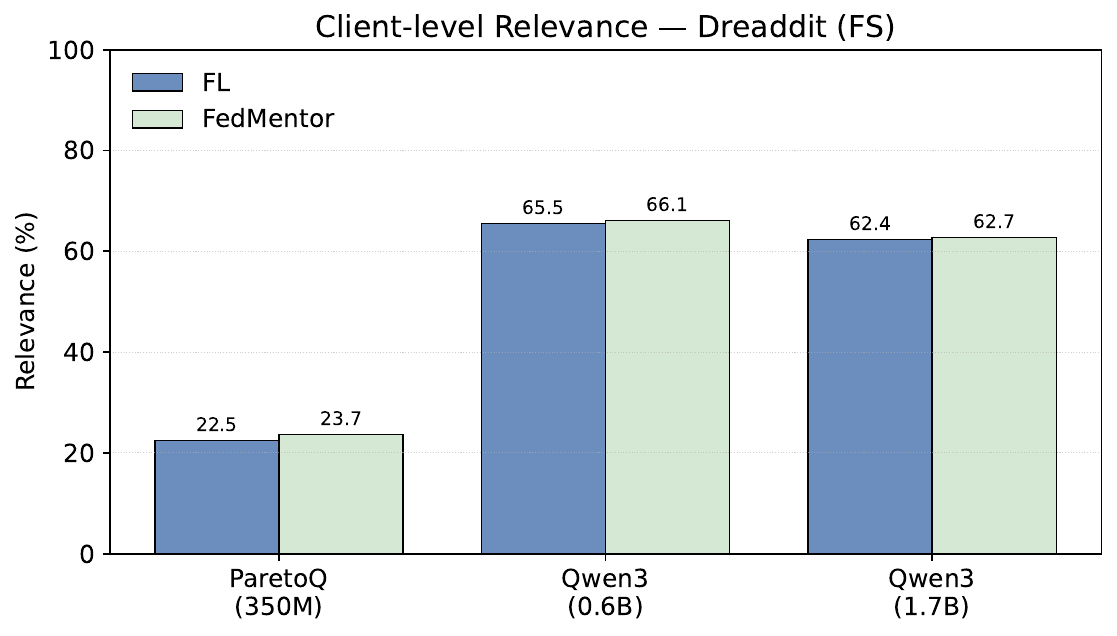}
    \caption{Dreaddit}
    \label{fig:rel_dreaddit}
  \end{subfigure}\hfill
  \begin{subfigure}[t]{0.33\textwidth}
    \centering
    \includegraphics[width=\linewidth]{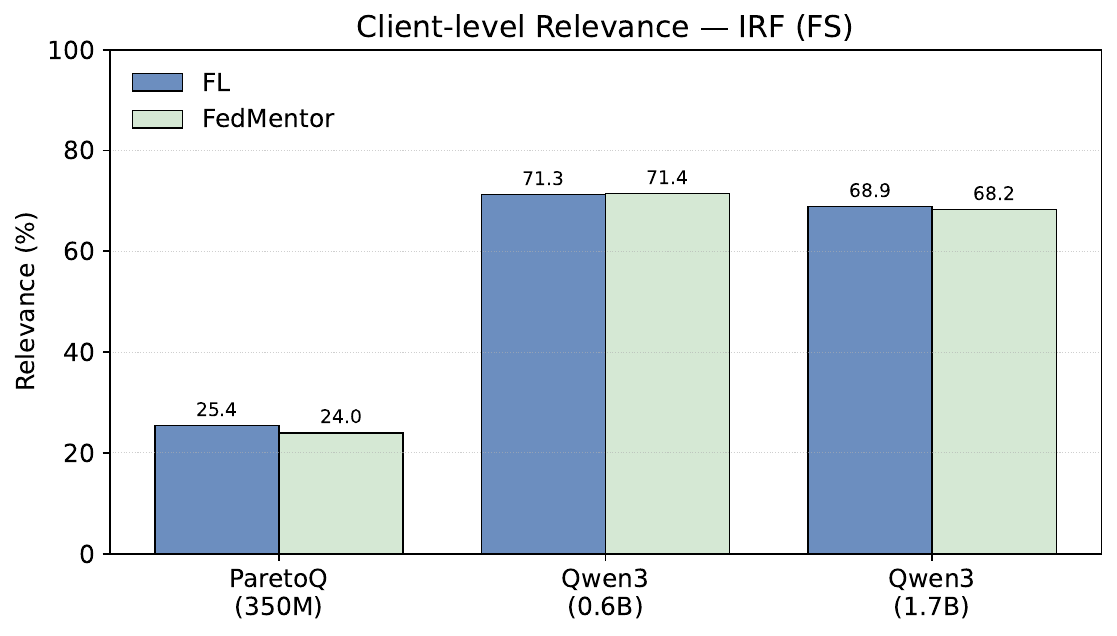}
    \caption{IRF}
    \label{fig:rel_irf}
  \end{subfigure}\hfill
  \begin{subfigure}[t]{0.33\textwidth}
    \centering
    \includegraphics[width=\linewidth]{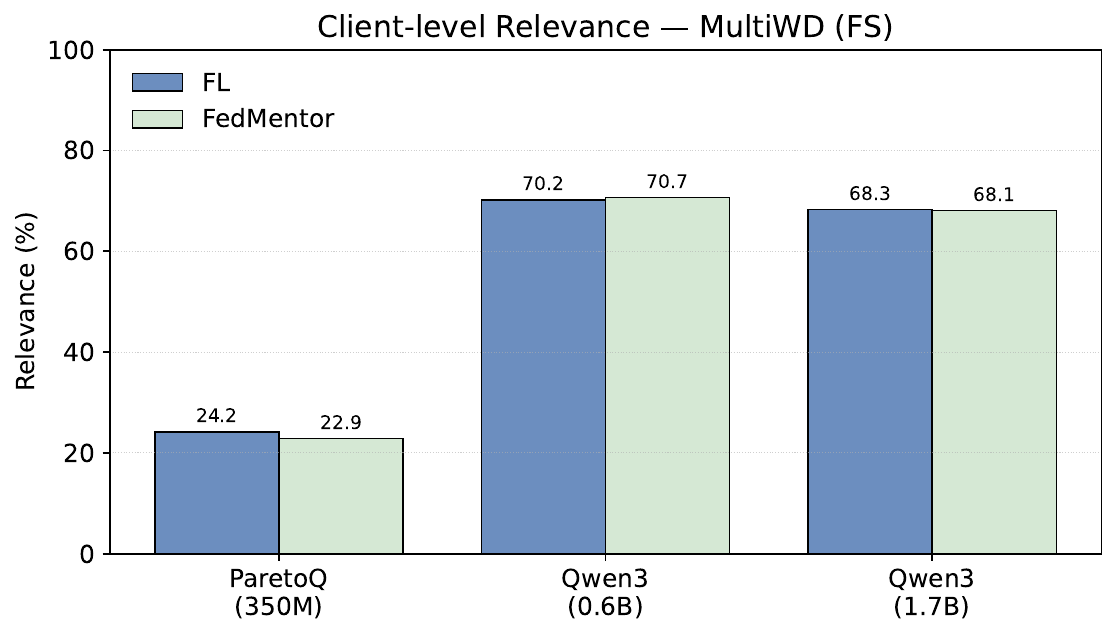}
    \caption{MultiWD}
    \label{fig:rel_multiwd}
  \end{subfigure}
  \caption{Client level relevance on three datasets. Bars compare FL and FedMentor for ParetoQ 350M, Qwen3 0.6B, and Qwen3 1.7B.}
  \label{fig:client_level_rel}
\end{figure*}

\subsection{Model Specification}
\label{app:models}

\noindent\textbf{MobileLLM-ParetoQ (350M)~\cite{liu2024mobilellm}.}
MobileLLM targets on device efficiency with a deep thin Transformer, embedding tying, grouped query attention, and blockwise weight sharing. ParetoQ adds sub 4 bit quantization with quantization aware training and learnable scaling. In \emph{FedMentor}, we load 4 bit NF4 with bf16 compute and fine-tune only LoRA adapters (rank 8), producing very small adapter states. Domain aware DP is applied directly to adapters, which keeps privacy cost and per round communication low.

\medskip
\noindent\textbf{SmolLM2 (360M and 1.7B)~\cite{allal2025smollm2}.}
SmolLM2 uses a data centric multi stage pipeline and efficient components such as RMSNorm, rotary embeddings, and gated feed forward layers. In \emph{FedMentor}, we freeze the backbone, attach LoRA adapters (rank 16), quantize to 4 bit NF4 with bf16 compute, and aggregate noised adapters with FedAvg. Utility-guided noise reduction maintains clinical metrics while preserving low-bandwidth through compact adapter updates.

\medskip
\noindent\textbf{Qwen3 (0.6B and 1.7B)~\cite{yang2025qwen3}.}
Qwen3 provides dense variants trained with instruction tuning and reinforcement learning for reasoning and remains memory efficient. In \emph{FedMentor}, Qwen3 integrates with our prompt manager for zero shot and few shot generation. We quantize the frozen backbone to 4 bit NF4 with bf16 compute, fine-tune LoRA adapters (rank 16), and apply domain specific DP budgets with layer and adapter aware noise scaling before FedAvg, achieving privacy preserving updates without logit sharing or knowledge distillation.

\subsection{Baseline Specification}
\label{app:baselines}

\noindent\textbf{Centralized (w/o FL, w/o DP).}
Pool all datasets and fine-tune a single LoRA adapter on the combined corpus without privacy constraints. Serves as an optimistic upper bound when central aggregation is hypothetically allowed.

\medskip

\noindent\textbf{Federated with LoRA (w/o DP).}
Standard FedAvg with LoRA adapters. Each client fine-tunes its local adapter while the backbone remains frozen. The server aggregates adapter parameters via data-weighted averaging, isolating the effect of decentralization.

\medskip

\noindent\textbf{Federated LoRA under Domain-Aware DP (FedMentor, Ours).} Each client applies domain specific privacy budgets to its LoRA parameters with layer and adapter aware noise scaling, assigning larger noise to more sensitive components and tighter budgets for higher risk domains. The server performs FedAvg over noised adapters, monitors clinical utility proxies during rounds, and applies utility guided noise reduction when thresholds are violated. This design preserves privacy, reduces communication by sharing only adapters, and stabilizes training under non IID client distributions.

\subsection{Detailed Evaluation Metrics}
\label{app:metrics}

We report metrics for quality, safety, and systems efficiency that match the implementation.

\noindent \textbf{Centralized (w/o FL, w/o DP).}  
We evaluate toxicity mean and toxicity max of generated responses using Detoxify, safe rate (fraction of responses below a fixed toxicity threshold), BERTScore F1 for semantic similarity, relevance mean via embedding based cosine similarity, ROUGE-L for lexical overlap, and perplexity mean as a fluency proxy. These quantify response quality and safety for single model generation.

\medskip

\noindent\textbf{Federated with LoRA (w/o DP).}  
For each client and in aggregate we report toxicity mean, safe rate, BERTScore F1, relevance mean, and ROUGE-L. In accordance with the code, toxicity max and perplexity mean are not computed in the federated evaluator.

\medskip

\noindent\textbf{Federated with LoRA (w/ DP) (System Metrics).}
To characterize privacy–utility tradeoffs and overhead, we log domain sensitivity, training time, final train loss, final eval loss, memory used (MB), and LoRA weights size (MB) per client, along with per round aggregates: average train loss, average eval loss, total training time, average memory usage, and communication overhead (MB). Communication overhead reflects the size of adapter payloads, since only LoRA adapters are shared rather than full model weights.

\begin{table*}[t]
\centering
\small
\setlength{\tabcolsep}{3pt}
\begin{tabular}{ll
  S[table-format=2.1] S[table-format=2.1]  % SR ZS/FS
  S[table-format=2.2] S[table-format=2.2]  % Tmx ZS/FS
  S[table-format=2.1] S[table-format=2.1]  % B-F1 ZS/FS
  S[table-format=2.2] S[table-format=2.2]  % R-L ZS/FS
  S[table-format=2.1] S[table-format=2.1]  % Rel ZS/FS
}
\toprule
\textbf{Setting} & \textbf{Method} &
\multicolumn{2}{c}{\textbf{SR (\%) $\uparrow$}} &
\multicolumn{2}{c}{\textbf{Tmx (\%) $\downarrow$}} &
\multicolumn{2}{c}{\textbf{B-F1 (\%) $\uparrow$}} &
\multicolumn{2}{c}{\textbf{R-L (\%) $\uparrow$}} &
\multicolumn{2}{c}{\textbf{Rel (\%) $\uparrow$}} \\
\cmidrule(lr){3-4}\cmidrule(lr){5-6}\cmidrule(lr){7-8}\cmidrule(lr){9-10}\cmidrule(lr){11-12}
& &
\textbf{ZS} & \textbf{FS} &
\textbf{ZS} & \textbf{FS} &
\textbf{ZS} & \textbf{FS} &
\textbf{ZS} & \textbf{FS} &
\textbf{ZS} & \textbf{FS} \\
\midrule
\multicolumn{12}{c}{\textbf{Dreaddit}}\\
\midrule
\multirow{2}{*}{Central}
& SmolLM2-360M & 98.0 & 98.0 & 76.44 & 79.45 & 82.6 & 83.3 & 6.35 & 7.79 & 23.5 & 28.1 \\
& SmolLM2-1.7B & 98.0 & 98.0 & 74.79 & 75.52 & 82.7 & 83.6 & 5.75 & 6.60 & 23.2 & 27.2 \\
\cmidrule(lr){1-12}
\multirow{2}{*}{\shortstack{FedMentor\\(FL w/ DP)}}
& SmolLM2-360M & 98.0 & 98.0 & 0.59 & 0.65 & 82.7 & 82.7 & 2.12 & 2.17 & 26.9 & 44.3 \\
& SmolLM2-1.7B & 97.0 & 98.0 & 0.54 & 0.68 & 82.2 & 82.0 & 2.02 & 2.49 & 31.6 & 0.5 \\
\midrule
\multicolumn{12}{c}{\textbf{IRF}}\\
\midrule
\multirow{2}{*}{Central}
& SmolLM2-360M & 98.0 & 98.0 & 36.00 & 62.42 & 84.0 & 86.3 & 2.52 & 11.85 & 27.1 & 27.1 \\
& SmolLM2-1.7B & 98.0 & 97.5 & 50.28 & 62.07 & 82.9 & 85.1 & 2.92 & 7.39 & 28.7 & 28.1 \\
\cmidrule(lr){1-12}
\multirow{2}{*}{\shortstack{FedMentor\\(FL w/ DP)}}
& SmolLM2-360M & 78.0 & 79.0 & 7.96 & 7.80 & 82.9 & 81.6 & 3.78 & 3.67 & 70.3 & 71.6 \\
& SmolLM2-1.7B & 77.0 & 79.0 & 9.09 & 8.23 & 83.0 & 81.3 & 3.38 & 3.55 & 67.1 & 0.0 \\
\midrule
\multicolumn{12}{c}{\textbf{MultiWD}}\\
\midrule
\multirow{2}{*}{Central}
& SmolLM2-360M & 97.5 & 98.5 & 47.74 & 61.66 & 84.4 & 86.7 & 6.23 & 11.57 & 39.8 & 26.8 \\
& SmolLM2-1.7B & 97.5 & 98.0 & 70.80 & 74.85 & 83.4 & 85.2 & 6.27 & 7.86 & 42.9 & 27.9 \\
\cmidrule(lr){1-12}
\multirow{2}{*}{\shortstack{FedMentor\\(FL w/ DP)}}
& SmolLM2-360M & 97.0 & 96.0 & 1.12 & 1.91 & 83.0 & 82.8 & 3.29 & 3.17 & 38.3 & 43.5 \\
& SmolLM2-1.7B & 97.0 & 96.0 & 1.59 & 2.43 & 81.7 & 78.9 & 2.99 & 3.64 & 41.8 & 0.1 \\
\bottomrule
\end{tabular}
\caption{Performance of SmolLM2 backbones under centralized and FedMentor (FL w/ DP) training on Dreaddit, IRF, and MultiWD. Columns report Zero-shot (ZS) and Few-shot (FS) for Safe Rate (SR), Toxicity max (Tmx), BERTScore F1 (B-F1), ROUGE L (R-L), and Relevance (Rel). All metrics are in \%.}
\label{tab:three-datasets-fit-2}
\end{table*}

\subsection{Prompt Construction for Zero-Shot and Few-Shot}
\label{app:prompt}

\textbf{Zero-Shot Prompting.}
We follow standard zero-shot evaluation, supplying a single instruction plus the post, wrapped in the backbone native template: Qwen uses chat role tags, SmolLM2 uses \textit{Input}/\textit{Response} headers, and MobileLLM uses plain text with a trailing \textit{Response:}. Instructions are domain aligned: stress identification and supportive coping for \textbf{Dreaddit}; empathetic interpersonal risk screening for \textbf{IRF} (thwarted belongingness or perceived burdensomeness); and brief practical guidance for social and emotional wellness cues in \textbf{MultiWD}. This uniform yet domain aware setup isolates instruction following while keeping comparisons consistent across datasets and models. Exact wrappers and examples are shown in Figure~\ref{fig:prompt_templates_zeroshot}, with dataset specific instructions in Figure~\ref{fig:dataset_specific_instructions}.

\medskip

\noindent\textbf{Few-Shot Prompting.}
We implement few-shot as supervised LoRA adaptation on each client rather than in context exemplars. Only adapters are trained on a quantized backbone; updates are privatized with domain aware Gaussian noise and aggregated with FedAvg. Short supportive targets align tone and structure to each domain, which increases safe rate, reduces toxicity mean, and improves BERTScore F1. At inference, we reuse the zero-shot wrapper for a clean comparison. The generic train time wrapper and target concatenation appear in Figure~\ref{fig:prompt_templates_fewshot}.

\subsection{Hardware Details} 
All experiments were conducted on a server with NVIDIA A100 GPUs, each with 80 GB of memory. Mixed-precision computation with bfloat16 (bf16) was used for both training and inference. Models were loaded in bf16 and executed with 4-bit NF4 quantization and bf16 compute. Up to two A100 GPUs were available, and each federated round allocated client training to a single GPU, enabling parallel execution across clients. This setup allowed efficient LoRA fine-tuning of models with up to 1.7B parameters and concurrent simulation of multiple clients. All training times, memory usage, and communication costs reported in this work were measured in this environment.

\section{Supplementary Results}
\label{app:Supplementaryresults}

Table~\ref{tab:three-datasets-fit-2} reports centralized and FedMentor performance for SmolLM2 on Dreaddit, IRF, and MultiWD. The pattern mirrors the main models: FedMentor keeps B-F1 and ROUGE close to baselines while improving safety.

% \noindent Table~\ref{tab:client_memory_gb_Qwen3_1.7B_IRF} presents per-round client memory usage for Qwen3-1.7B under the IRF $\epsilon$-sensitivity test with $\epsilon=1.0$. The sweep varies only the IRF domain privacy budget, while Dreaddit and MultiWD are held fixed at 2.0 and 1.5, respectively. These results complement the main experiments by showing how memory demands evolve across training rounds when privacy constraints differ across domains.

\noindent Figure~\ref{fig:client_level_rel} compares FL and FedMentor for ParetoQ 350M, Qwen3 0.6B, and Qwen3 1.7B; per-client relevance remains aligned, indicating stable utility under non-IID data and no domain collapse.

\section{Use of AI-Assisted Tools}
During development, we used GitHub Copilot to suggest code completions and refactorings. For writing, we utilized ChatGPT to proofread and refine phrasing. All inputs to these tools originated with the authors, and every suggestion was reviewed and edited by the authors, who remain responsible for the final content. 

\end{document}